\documentclass[aps,prc,twocolumn,nofootinbib,floatfix,amsmath,amssymb,amsfonts,superscriptaddress]{revtex4-2}

\usepackage[T1]{fontenc}
\usepackage[utf8]{inputenc}
\usepackage[english]{babel}

\usepackage{times}
\usepackage{txfonts}
\usepackage{graphicx} 
\usepackage{dcolumn}  
\usepackage{xcolor}

\usepackage[range-units=single]{siunitx}
\DeclareSIUnit{\fm}{\femto\meter}
\DeclareSIUnit{\fmfour}{\femto\meter\tothe{4}}
\DeclareSIUnit[quantity-product = {}]{\percent}{\%}

\usepackage{xspace}

\usepackage{dcolumn}
\newcolumntype{d}[1]{D{.}{.}{#1}}

\usepackage[
   colorlinks=true,
   citecolor=blue,
   linkcolor=blue,
   urlcolor=blue,
]{hyperref}
\usepackage{orcidlink}
\usepackage[]{cleveref}

\crefname{table}{Table}{Tables}
\Crefname{table}{Table}{Tables}
\crefname{figure}{Fig.}{Figs.}
\Crefname{figure}{Figure}{Figures}
\crefname{section}{Sec.}{Secs.}
\Crefname{section}{Section}{Sections}
\crefname{equation}{Eq.}{Eqs.}
\Crefname{equation}{Equation}{Equations}

\DeclareMathAlphabet{\curly}{OMS}{cmsy}{m}{n}

\newcommand{\Rprms}{\ensuremath{R_p}\xspace}
\newcommand{\Rnrms}{\ensuremath{R_n}\xspace}

\newcommand{\DRpn}{\ensuremath{\Delta R_{p,n}}\xspace}
\newcommand{\DSRprms}{\ensuremath{\Delta R^2_p}\xspace}
\newcommand{\DSRch}{\ensuremath{\Delta R^2_\mathrm{ch}}\xspace}

\newcommand*{\elem}[2]{\ensuremath{{}^\text{#2}\text{#1}}\xspace}

\newcommand{\B}[1]{\elem{B}{#1}}
\newcommand{\Li}[1]{\elem{Li}{#1}}
\newcommand{\He}[1]{\elem{He}{#1}}

\newcommand{\Nmax}{\ensuremath{N_\text{max}}\xspace}
\newcommand*{\cNmax}{\ensuremath{{\curly{N}_\mathrm{max}}}\xspace}

\newcommand{\NxLO}[2][]{
    \ifcase#2
        \ensuremath{\text{LO}^{#1}}
    \or
        \ensuremath{\text{NLO}^{#1}}
    \else
        \ensuremath{\text{N}^{#2}\text{LO}^{#1}}
    \fi
}

\newcommand{\aHO}{\ensuremath{a_\text{HO}}\xspace}

\begin{document}

\title{High-Precision \emph{Ab Initio} Radius Calculations of Boron Isotopes}

\author{T.~Wolfgruber\,\orcidlink{0000-0003-0933-0682}}
\email{tobias.wolfgruber@physik.tu-darmstadt.de}
\affiliation{Institut f\"ur Kernphysik, Technische Universit\"at Darmstadt, 64289 Darmstadt, Germany}
\author{T.~Gesser\,\orcidlink{0009-0006-5401-7045}}
\affiliation{Institut f\"ur Kernphysik, Technische Universit\"at Darmstadt, 64289 Darmstadt, Germany}
\author{M.~Kn\"oll\,\orcidlink{0009-0001-6023-8657}}
\affiliation{Institut f\"ur Kernphysik, Technische Universit\"at Darmstadt, 64289 Darmstadt, Germany}
\author{P.~Maris\,\orcidlink{0000-0002-1351-7098}}
\affiliation{Dept.\ of Physics and Astronomy, Iowa State University, Ames, Iowa 50011, USA}
\author{R.~Roth\,\orcidlink{0000-0003-4991-712X}}
\email{robert.roth@physik.tu-darmstadt.de}
\affiliation{Institut f\"ur Kernphysik, Technische Universit\"at Darmstadt, 64289 Darmstadt, Germany}
\affiliation{Helmholtz Forschungsakademie Hessen f\"ur FAIR, GSI Helmholtzzentrum, 64289 Darmstadt, Germany}

\date{\today}

\begin{abstract}
We perform a precision study of radii in Boron isotopes for multiple realistic interactions from chiral effective field theory.
We obtain predictions of radii with combined many-body and interaction uncertainty quantification from \emph{ab initio} no-core shell model calculations together with machine learning extrapolation methods.
An extension to radius differences further allows us to investigate a potential proton halo in \B{8} and, moreover, provide predictions that relate directly to the isotope shift, which can be precisely measured in experiments.
\end{abstract}

\maketitle

\section{Introduction}

Radii and ground-state energies are fundamental observables in nuclear structure physics, providing insights into the strong interaction between nucleons.
These quantities are crucial for understanding the size, structure, and spatial distribution of matter within atomic nuclei.
In this context, the term ``radii'' collectively refers to the charge radius, which can be measured experimentally, and the point-proton and point-neutron rms radius, which are the result of calculations treating nucleons as point-like particles.
Accurate theoretical predictions are crucial for advancing our understanding of nuclei, particularly for light elements such as boron isotopes, which can be calculated with higher precision compared to heavier nuclei.
The charge radii of the stable Boron isotopes \B{10} and \B{11} have been extensively studied, with theoretical predictions validated against high-precision experimental measurements \cite{Ber19}.
Beyond the stable isotopes, neutron-deficient nuclei like \B{8} are of great interest due to their potential for exhibiting a proton halo.
This makes \B{8} an ideal candidate for charge radius investigations, as the charge radius reflects the proton distribution.
While some experimental data, such as quadrupole moment measurements, suggest a halo character \cite{Sumikama:2006av}, theoretical predictions of the charge radius remain essential for understanding this phenomenon.
Future high-precision experimental measurements of the \B{8} charge radius will provide additional data to further test theoretical models \cite{Ber17}.

Among these theoretical methods, the no-core shell model (NCSM) \cite{BaNa13,NaQu09,Roth09,ZheBa93} is one of the most commonly used \emph{ab initio} method for calculating nuclear observables in light nuclei based on realistic interactions from chiral effective field theory (EFT) \cite{Epelbaum2009Modern,Machleidt2011Chiral}.
While it gives access to the full suite of nuclear structure observables, the reach and precision of the NCSM are limited by rapidly growing model-space dimensions, especially with the number of nucleons.
Radii, specifically, require large model spaces due to their sensitivity to the long-range part of the many-body wavefunction.
Hence, sophisticated and reliable extrapolation methods are required to extract precise radii, with quantified many-body truncation uncertainties, from NCSM calculations in finite model spaces.

Recently, machine learning tools have been introduced as complemental methods for overcoming the computational limitations of \emph{ab initio} calculations \cite{NeVa19,JiHa19,KnoWo23,WoKno24,mazur2024machine,Knoell:2025otn,Knoell:2025bm}.
Previous studies have demonstrated the effectiveness of artificial neural networks (ANNs) in predicting converged values of nuclear observables, in particular energies, radii, and electric quadrupole moments, from NCSM calculations in small model spaces, including statistical estimates for the uncertainties associated with the model-space truncation \cite{KnoWo23,WoKno24,Knoell:2025otn}.
In addition, interaction uncertainties resulting from the omission of higher orders in the chiral expansion can be quantified via Bayesian analysis \cite{MeFu19}.

In this work, we employ the NCSM together with ANN extrapolations for a precision study of radii in Boron isotopes.
We present predictions for point-proton (\Rprms) and point-neutron rms radii (\Rnrms) with combined many-body and interaction uncertainties and compare results for two different families of chiral interactions.
We further extend the extrapolation scheme to differences of these radii $\DRpn=R_p-R_n$, which allow us to investigate the aforementioned halo structures.
Moreover, the difference of the squared point-proton radii of different isotopes, $\DSRprms = \DSRch + \left(\frac{N_1}{Z_1}-\frac{N_2}{Z_2}\right)r^2_n$,
with proton and neutron numbers $N_i$, $Z_i$ and neutron charge radius $r_n$ is of particular interest, as it directly relates to the difference of the charge radii \DSRch, which is accessible in isotope-shift measurements with high precision \cite{Ber19}.

\section{Methods}

\subsection{No-Core Shell Model}

In the NCSM all nucleons are considered active degrees of freedom and the many-body problem is expanded in a basis of Slater determinants, constructed from a harmonic oscillator (HO) single-particle basis, converting the Schr\"odinger equation into a matrix eigenvalue problem.
To ensure computational feasibility, the infinite-dimensional Hilbert space is truncated, defining a finite model space controlled by the truncation parameter \Nmax.
This parameter limits the number of excitation quanta above the lowest-energy determinant, such that the solution in the full Hilbert space is recovered in the limit $\Nmax\to\infty$.
Together with the length scale of the underlying HO basis, the oscillator length \aHO, the truncation parameter defines the convergence pattern of an observable in a sequence of growing model spaces.
By construction, the sequences for different \aHO eventually converge towards the same value for sufficiently large \Nmax.
The convergence rate is accelerated through a similarity renormalization group (SRG) transformation of the Hamiltonian prior to the NCSM calculation \cite{Bogner2007Similarity,Bogner:2007rx,Roth2014evolved}.
For energies the NCSM is variational, hence, a monotonously decreasing convergence emerges.
This, however, does not apply to radii, resulting in ascending, descending, and oscillating sequences with a particularly strong dependence on \aHO.
While complete convergence can be achieved for few-body systems such as \He{4}, p-shell nuclei already pose a considerable computational challenge.
This results in insufficient convergence, which necessitates the use of extrapolation schemes.

\subsection{Artificial Neural Network Extrapolations}

ANNs have proven to be effective tools for addressing the extrapolation challenge, utilizing their pattern recognition capabilities to predict converged values from NCSM calculations.
We employ the ANNs developed in Ref.~\cite{WoKno24}, i.e., fully-connected feed-forward neural networks with 12 input nodes, hidden layers of sizes 48, 48, and 24, and a single output node using ReLU as the activation function.
These ANNs are trained on an extensive library of NCSM data for few-body systems ($A\leq 4$) and yield predictions for converged energies and radii in p-shell nuclei.
A single prediction is based on a sample of NCSM results for 3 values of \aHO at 4 consecutive \Nmax, so for example 12 radius values, one for each \aHO and \Nmax combination.
The sample is then normalized and fed into the ANN, which outputs its prediction for the fully converged value, from which the converged observable can be recovered by inverting the normalization.
By using 1000 ANNs and constructing all possible input samples from the evaluation data one obtains a distribution of predictions that allows for the extraction of uncertainties via statistical means.
In order to do so, we first depict the predictions as a histogram using the Freedman-Diaconis rule~\cite{FrDi81}.
The histogram is then smoothed using a running mean of widths 10, and the maximum of the distribution, the most probable value, defines the nominal prediction.
To estimate a confidence interval, bins to the left and right of the most probable value are added up, always continuing with the larger one, until \qty{68}{\percent} of the predictions are included.
This results in an asymmetric interval that resembles a 1-$\sigma$ uncertainty and includes both many-body and network-related contributions (see Ref.~\cite{Knoell:2025otn} for details).

When evaluating the ANNs with NCSM sequences for a wide range of \aHO, we find a disproportionate growth of the uncertainty, while the most probable value barely shifts.
Additionally, the inclusion of \aHO far from the optimum can introduce unwanted artifacts, such as multimodality, in the histograms.
In order to mitigate these effects, we introduce a pre-selection of HO lengths that is implemented before the extrapolation in order to enhance the precision of the predictions.
In particular, we restrict the evaluation data for energies to the 5 oscillator lengths that yield the lowest-lying energies at the highest \Nmax and for radii
we select the 4 flattest ascending sequences available.
\begin{figure*}
    \centering
    \includegraphics[width=\linewidth]{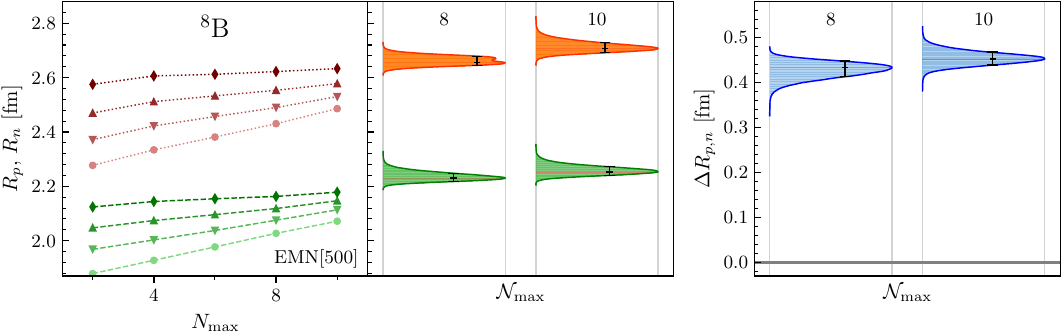}
    \caption{Input data and network predictions for the point-proton (orange) and point-neutron (green) radius of \B{8} with the resulting radius difference for the EMN[500] interaction. In the left panel are the NCSM calculations for both radii, with their corresponding histograms of the predictions of the networks. The selection of \aHO was done individually for each radius, with $\aHO(\Rprms)$ ranging from 1.6 to \qty{1.9}{\fm} and $\aHO(\Rnrms)$ from 1.4 to \qty{1.7}{\fm}. The slight fragmentation around the maximum of the $\Rprms$ histogram at $\cNmax{}=8$ can be remedied through further optimization of the \aHO{} selection. The right panel shows the calculated radius difference $\DRpn$, obtained by subtracting the network predictions of the radii samplewise.}
    \label{fig:B8_Rp_Rn_DRpn}
\end{figure*}
This ensures precise predictions for both observables with reasonable uncertainties as shown in the left-hand and center panels of \cref{fig:B8_Rp_Rn_DRpn}.
Depicted are the selected NCSM sequences for \Rprms and \Rnrms for \B{8} along with the resulting histograms of distributions grouped by \cNmax, i.e., the highest \Nmax present in a given sample.
We find stable predictions for both radii with large overlap, and therefore good agreement, between the histograms at different \cNmax.
For a more comprehensive overview of the ANNs performance and comparisons with other extrapolation methods, see Ref.~\cite{Knoell:2025bm}.


The ANNs can further be employed to obtain predictions for the differences of radii.
This is done in analogy to the extrapolations of excitation energies and hyperon separation energies in Refs.~\cite{WoKno24,Knoell2023Hyperon}.
Each prediction is associated with a specific input sample.
Instead of extrapolating both radii individually and propagating the corresponding uncertainties, we resort to a samplewise subtraction, generating a new distribution prior to the statistical analysis.
This results in a new histogram for the radius differences, where the correlations between the two observables are inherently included, which results in notably reduced uncertainties.
While this is straight-forward for energy differences, where one can, in general, subtract samples with the same sets of \aHO and \Nmax, this is more complicated for radius differences.
Due to the strong correlation between radius and \aHO, both radii often have different optimal ranges of \aHO.
In order to account for this, we find the optimal \aHO range for both radii independently, and subtract the predictions for the respective samples of each radius from each other.
This leaves us with a distribution for the radius differences, which, again, exhibits a significantly reduced uncertainty.
An example for such an analysis of \DRpn in \B{8} is shown in the right-hand panel of \cref{fig:B8_Rp_Rn_DRpn}.
Analogously this can be applied to radius difference between isotopes.
The only difference is that, instead of directly subtracting the samples, we square the values of each sample before subtracting.
\begin{figure*}
    \centering
    \includegraphics[width=\linewidth]{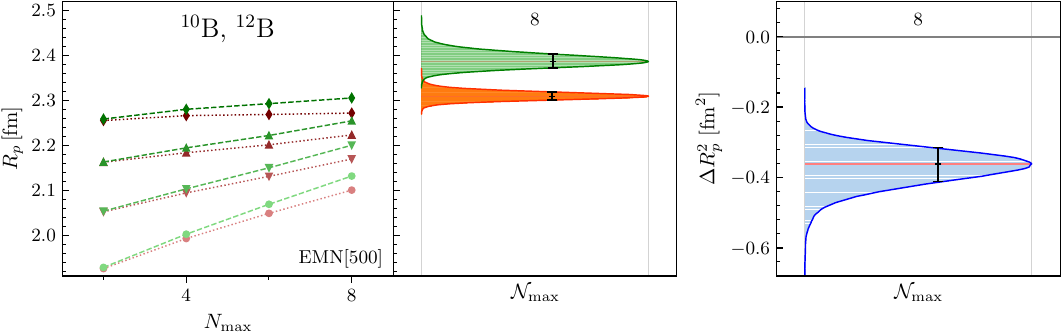}
    \caption{Input data and network predictions for the point-proton radius of \B{10} (green) and \B{12} (orange) and the resulting squared radius difference for the EMN[500] interaction. The left panel shows the input data consisting of NCSM calculations for both nuclei with $a_\text{HO}$ ranging from 1.3 to \qty{1.6}{\fm}, along with corresponding histograms for the network predictions. On the right is the difference of the squared proton radius $\DSRprms$, calculated by subtracting the squared values of the network predictions samplewise, resulting in a new distribution.}
    \label{fig:B10B11_Rp_DR2p}
\end{figure*}
An example for \DSRprms between \B{10} and \B{12} is given in \cref{fig:B10B11_Rp_DR2p}, where the left-hand and center panels again show the selected NCSM data and the corresponding ANN predictions for the respective \Rprms, while the right-hand panel depicts the extracted difference.


\subsection{Interaction Uncertainties}

So far, we have limited our discussion of uncertainties to those arising from the model-space truncation.
However, an additional, and often dominant, source of uncertainty is the truncation of the chiral expansion resulting in an interaction uncertainty that needs to be estimated.
Chiral interactions are organized in different orders and the order-by-order behavior can be exploited to estimate the missing contributions of higher orders beyond the truncation.
A recipe for the estimation of truncation errors in chiral EFT has been developed by the BUQEYE collaboration \cite{MeFu19}.
We perform NCSM calculations at each chiral order available and employ the BUQEYE pointwise model based on the most probable values at each order.
This way, we obtain a Student-$t$ distribution that accounts for the interaction uncertainties.
In order to incorporate this into the distribution of predictions obtained from our ANNs, the Student-$t$ distribution is folded with the distribution of ANN predictions, resulting in a broadened distribution that includes both many-body and interaction uncertainties, while the most probable value remains nearly the same.
From this new distribution we can, again, statistically extract our final result.
This procedure is illustrated in \cref{fig:chiral_EMN500,fig:chiral_SMS500}, which we will discuss in detail in the next section.

Note that we do not explicitly include the uncertainties arising from the choice of regulator scheme or cutoff scale into the uncertainty estimate. However, we illustrate and discuss them by comparing calculations for different interaction families and cutoffs.

\section{Results}

For an extensive study of radii in Boron isotopes we employ two different families of NN+3N interactions from chiral EFT; a nonlocal interaction by Entem, Machleidt, and Nosyk (EMN) \cite{EnMa17} supplemented with nonlocal 3N interactions up to N${}^3$LO \cite{HuVo20}, and a semilocal momentum space interaction (SMS) by the LENPIC collaboration \cite{LENPIC21,Mar22}.
For each family we use three chiral orders with cutoffs of 450 and \qty{500}{\MeV}, respectively.
All interactions are SRG evolved to $\alpha=\qty{0.08}{\fmfour}$, and SRG corrections to the radius operator are included up to the two-body level.
The NCSM calculations are performed on an \aHO grid from 1.2 to \qty{2.4}{\fm} in steps of \qty{0.1}{\fm}.

\begin{figure}
    \centering
    \includegraphics[width=\linewidth]{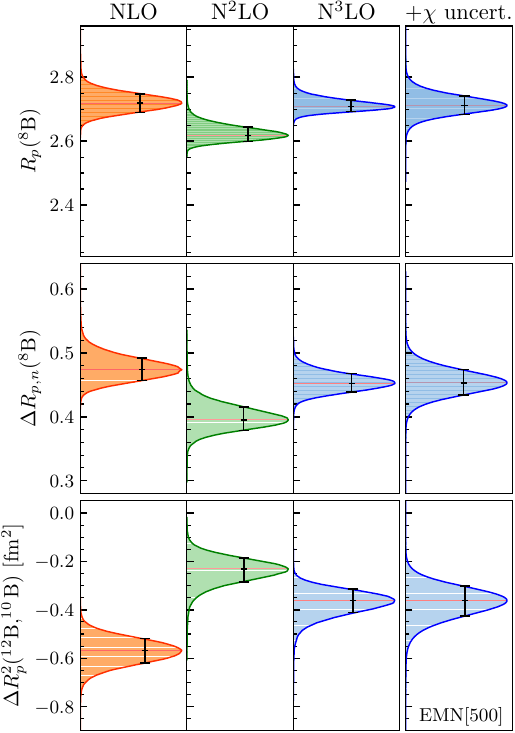}
    \caption{Network predictions for radii of boron isotopes from NCSM calculations using different orders of the nonlocal interaction family (EMN) with a cutoff of $\Lambda = \qty{500}{\MeV}$. The first three columns of each panel show histograms for \NxLO{1}, \NxLO{2} and \NxLO{3}, while the histograms in the last column combine the \NxLO{3} results with chiral uncertainties, estimated via Bayesian analysis. The first panel shows the ANN predictions for the point-proton radius of \B{8}, while in the second panel the differences between the point-proton and point-neutron radius are displayed. In the last panel are the squared point-proton radius differences of \B{10} and \B{12}.}
    \label{fig:chiral_EMN500}
\end{figure}
\begin{figure}
    \centering
    \includegraphics[width=\linewidth]{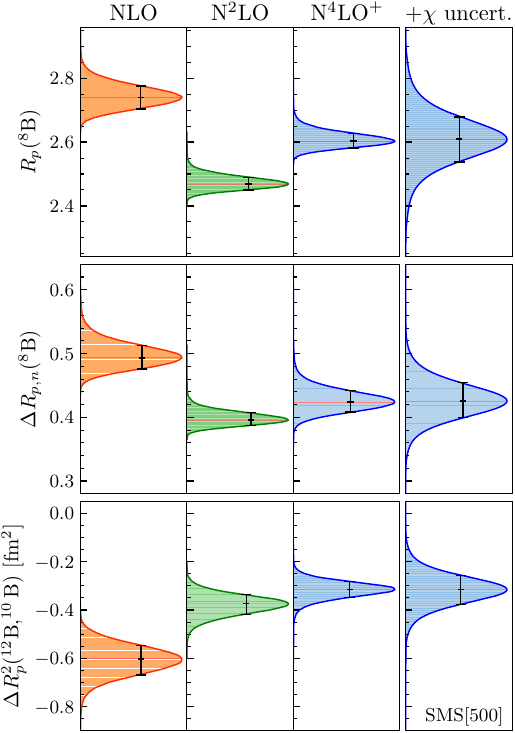}
    \caption{Network predictions for radii of boron isotopes from NCSM calculations using different orders of the semilocal interaction family (SMS) with a cutoff of $\Lambda = \qty{500}{\MeV}$. The first three columns of each panel show histograms for \NxLO{1}, \NxLO{2} and \NxLO[+]{4}, while the histograms in the last column combine the \NxLO[+]{4} results with chiral uncertainties, estimated via Bayesian analysis. The first panel shows the predictions for the point-proton radius of \B{8}, while in the second panel the differences between the point-proton and point-neutron radius are displayed. In the last panel are the squared point-proton radius differences of \B{10} and \B{12}.}
    \label{fig:chiral_SMS500}
\end{figure}

We start our investigations with the order-by-order behavior of \Rprms and \DRpn in \B{8} along with \DSRprms between \B{10} and \B{12} for both interaction families, as depicted in \cref{fig:chiral_EMN500,fig:chiral_SMS500}, respectively.
Histograms of ANN predictions are shown for each chiral order, and the rightmost column contains histograms for the highest order that additionally include chiral uncertainties.
For the EMN interaction NN+3N forces are consistently available at next-to-leading order ($\NxLO{1}$), next-to-next-to-leading order ($\NxLO{2}$), and $\NxLO{3}$.
For the SMS interaction, this only holds for $\NxLO{1}$ and $\NxLO{2}$ since the 3N forces are not available at $\NxLO{3}$.
Hence, $\NxLO[+]{4}$ only contains $\NxLO{2}$ three-body forces and we, therefore, regard it as another third-order result for the BUQEYE analysis.
This naturally results in larger interaction uncertainties for the SMS interaction.
Comparing both figures for \Rprms of \B{8} we find that the $\NxLO{2}$ results exhibit a shift to smaller radii, which is then remedied in the next order.
This shift is significantly more pronounced for the SMS interaction.
This translates to \DRpn, which shows a similar behavior as \Rprms, yet the size of the shift is now about the same for both interactions.
When it comes to the radius difference between different isotopes, the order-by-order behavior is less systematic and strongly depends on the interaction and the two isotopes involved.
In the depicted example for \B{10} and \B{12} in the lower panels of \cref{fig:chiral_EMN500,fig:chiral_SMS500} we find that the $\NxLO{1}$ results tend to overestimate the absolute radius difference, resulting in a trend to smaller values with increasing chiral order.
In general, we find a very similar order-by-order behavior for both families of interactions across the different observables, and note that the chiral uncertainties scale with the changes between the different orders, as expected.

\begin{figure}
    \centering
    \includegraphics[width=\linewidth]{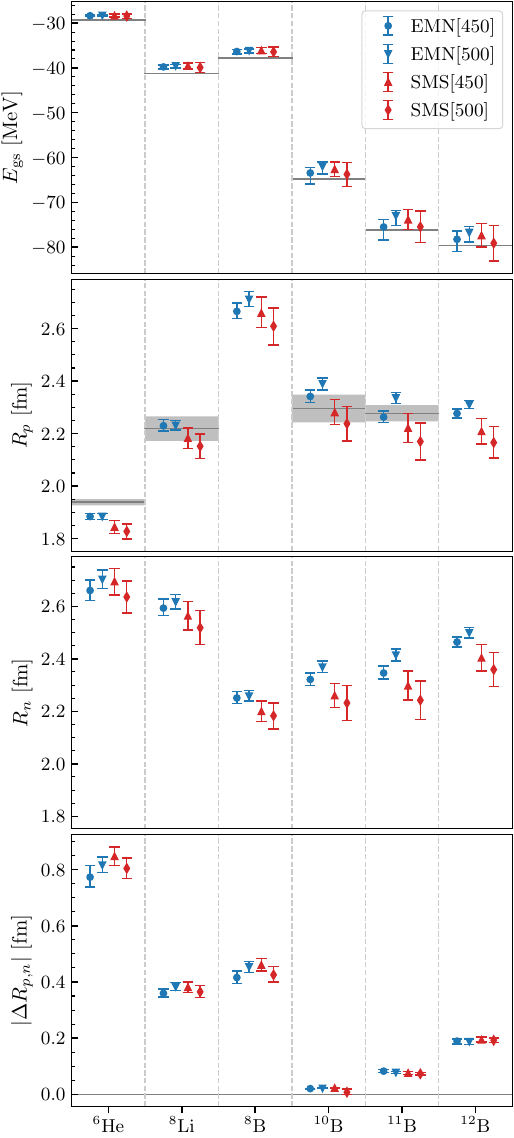}
    \caption{Predicted ground-state energies, radii and radius differences for \He{6}, \Li{8} and boron isotopes. Each column displays the predicted values of the corresponding observable of the respective nucleus, calculated at the highest order for each interaction family and cutoff. The predictions include chiral uncertainties. The first two panels show the results for the ground-state energy and the point-proton radius \Rprms. Gray bands indicate experimental values taken from Refs.~\cite{Wang:2021xhn,Angeli:2013epw}. Similarly, the second panel presents the results for the point-neutron radii \Rnrms. The final panel shows the absolute value of the differences of the predicted point-proton and point-neutron radii \DRpn.}
    \label{fig:Rp_Rn_AbsDRpn_overview}
\end{figure}

Next, we extend our investigations to a more general study of the Boron isotopes \B{8,10,11,12} for the EMN and SMS interactions with cutoffs $\Lambda=450$ and \qty{500}{\MeV}.
As we focus on \B{8} in particular, we further include its mirror nucleus \Li{8} and additionally consider \He{6} as an accepted halo nucleus \cite{Tanihata:2013jwa} for comparison.
\Cref{fig:Rp_Rn_AbsDRpn_overview} shows the results for the ground-state energy, \Rprms, \Rnrms, and $|\DRpn|$ obtained with all four interactions at the highest order, respectively.
Error bars indicate combined many-body and interaction uncertainties and experimental values are given in gray where available.
First, we find that the ground-state energies, shown in the top panel, are in very good agreement across interactions and overall agree well with experiment, albeit with a slight tendency to underbinding.
The correct reproduction of the energies warrants that features observed in the radii are not merely consequences of deviations in the binding energy.
This leads us to \Rprms and \Rnrms depicted in the center panels.
Here, we find an enhanced dependence on the interaction, where the SMS interactions generally tend to predict smaller radii, still, both interactions at $\Lambda=\qty{450}{\MeV}$ are mostly in agreement within uncertainties.
It is interesting to note that the change of radii with cutoff seems to be in opposite directions for EMN and SMS.
In comparison with experiment we find that the \Rprms agree within uncertainties except for \He{6}, where it is underestimated.
For \Rnrms there is no experimental data available, however, we can qualitatively assess \DRpn in order to investigate halo structures.

Focusing on the proton-halo candidate \B{8}, we find that \Rprms significantly exceeds \Rnrms leading to a much larger radius difference for \B{8} compared to the other boron isotopes, where the increasing radius difference is primarily attributed to the growing number of neutrons as evident from the bottom panel in \cref{fig:Rp_Rn_AbsDRpn_overview}.
This is mirrored in \Li{8}, where an analogous neutron excess leads to a radius difference of similar magnitude.
The remaining difference can likely be attributed to the additional Coulomb repulsion of the protons.
For both \Li{8} and \B{8}, however, the difference between proton and neutron radius is not as pronounced as in \He{6}.
Note that the radius differences are in much better agreement across the interactions than the individual radii and come with significantly smaller uncertainties.

\begin{figure}
    \centering
    \includegraphics[width=\linewidth]{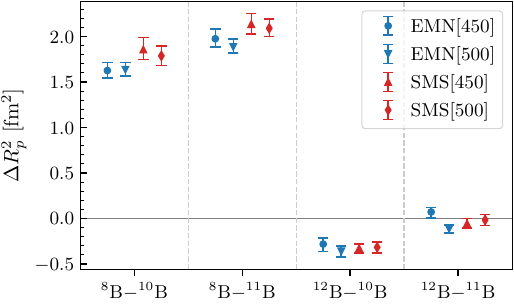}
    \caption{Squared point-proton radius differences for \B{8} and \B{12}, using \B{10} and \B{11} as a reference, calculated from NCSM data using various interaction families and cutoffs. Each column presents the isotope differences for all interactions used, with values obtained from the highest order of the respective interaction family, including chiral uncertainties. Positive (negative) values indicate that the radius is larger (smaller) than that of the \B{10} or \B{11} reference nucleus.}
    \label{fig:DSCRprms}
\end{figure}

\begin{table*}
    \centering
    \renewcommand{\arraystretch}{1.25}
    \begin{ruledtabular}
        \begin{tabular}{@{}l@{\extracolsep{13pt}}  d{2.8}@{\extracolsep{8pt}} d{2.8} d{2.8} d{2.7}@{\extracolsep{20pt}}   d{2.8}@{\extracolsep{8pt}} d{2.8} d{2.8} d{2.7}@{}}
        & \multicolumn{4}{c}{nonlocal (EMN) $\Lambda=\qty{450}{\MeV}$} & \multicolumn{4}{c}{nonlocal (EMN) $\Lambda=\qty{500}{\MeV}$}\\
        \cline{2-5}\cline{6-9}\\[-1.3em]
        Observable  &  \multicolumn{1}{c}{\NxLO{1}}  &  \multicolumn{1}{c}{\NxLO{2}}  &  \multicolumn{1}{c}{\NxLO{3}}  &  \multicolumn{1}{c}{$+\chi$ uncert.} & \multicolumn{1}{c}{\NxLO{1}}  &  \multicolumn{1}{c}{\NxLO{2}}  &  \multicolumn{1}{c}{\NxLO{3}}  &  \multicolumn{1}{c}{$+\chi$ uncert.}\\%
        \hline
    \Rprms(\He{6})  &  1.843_{-0.008}^{+0.008}  &  1.876_{-0.008}^{+0.010}  &  1.883_{-0.010}^{+0.010}  &  1.884_{-0.012}^{+0.012}  &  1.885_{-0.009}^{+0.008}  &  1.864_{-0.009}^{+0.009}  &  1.883_{-0.009}^{+0.009}  &  1.884_{-0.011}^{+0.013}\\
\Rprms(\Li{8})  &  2.161_{-0.014}^{+0.011}  &  2.211_{-0.013}^{+0.021}  &  2.228_{-0.016}^{+0.021}  &  2.230_{-0.020}^{+0.023}  &  2.208_{-0.015}^{+0.012}  &  2.202_{-0.014}^{+0.018}  &  2.229_{-0.013}^{+0.017}  &  2.231_{-0.017}^{+0.019}\\
\Rprms(\B{8})  &  2.620_{-0.028}^{+0.031}  &  2.618_{-0.025}^{+0.032}  &  2.663_{-0.023}^{+0.030}  &  2.666_{-0.028}^{+0.033}  &  2.719_{-0.029}^{+0.029}  &  2.618_{-0.019}^{+0.026}  &  2.708_{-0.016}^{+0.021}  &  2.711_{-0.027}^{+0.030}\\
\Rprms(\B{10})  &  2.257_{-0.016}^{+0.019}  &  2.310_{-0.015}^{+0.020}  &  2.340_{-0.019}^{+0.021}  &  2.342_{-0.023}^{+0.025}  &  2.344_{-0.015}^{+0.017}  &  2.319_{-0.010}^{+0.015}  &  2.386_{-0.015}^{+0.017}  &  2.389_{-0.023}^{+0.023}\\
\Rprms(\B{11})  &  2.208_{-0.011}^{+0.014}  &  2.237_{-0.020}^{+0.015}  &  2.261_{-0.016}^{+0.021}  &  2.263_{-0.020}^{+0.023}  &  2.287_{-0.013}^{+0.015}  &  2.262_{-0.011}^{+0.016}  &  2.332_{-0.010}^{+0.015}  &  2.335_{-0.020}^{+0.022}\\
\Rprms(\B{12})  &  2.142_{-0.010}^{+0.013}  &  2.271_{-0.012}^{+0.009}  &  2.276_{-0.011}^{+0.012}  &  2.277_{-0.016}^{+0.018}  &  2.219_{-0.011}^{+0.013}  &  2.270_{-0.010}^{+0.012}  &  2.309_{-0.008}^{+0.010}  &  2.310_{-0.015}^{+0.016}\\[0.4em]
\DRpn(\He{6})  &  -0.863_{-0.021}^{+0.021}  &  -0.737_{-0.040}^{+0.031}  &  -0.773_{-0.042}^{+0.032}  &  -0.773_{-0.042}^{+0.036}  &  -0.866_{-0.023}^{+0.023}  &  -0.763_{-0.027}^{+0.020}  &  -0.816_{-0.027}^{+0.022}  &  -0.816_{-0.029}^{+0.027}\\
\DRpn(\Li{8})  &  -0.370_{-0.014}^{+0.017}  &  -0.325_{-0.014}^{+0.009}  &  -0.358_{-0.014}^{+0.011}  &  -0.359_{-0.016}^{+0.014}  &  -0.383_{-0.018}^{+0.014}  &  -0.336_{-0.013}^{+0.008}  &  -0.381_{-0.011}^{+0.008}  &  -0.382_{-0.016}^{+0.012}\\
\DRpn(\B{8})  &  0.448_{-0.019}^{+0.020}  &  0.389_{-0.012}^{+0.015}  &  0.416_{-0.021}^{+0.021}  &  0.416_{-0.021}^{+0.022}  &  0.474_{-0.017}^{+0.018}  &  0.395_{-0.017}^{+0.020}  &  0.452_{-0.013}^{+0.016}  &  0.453_{-0.019}^{+0.021}\\
\DRpn(\B{10})  &  0.021_{-0.001}^{+0.001}  &  0.019_{-0.001}^{+0.001}  &  0.020_{-0.001}^{+0.001}  &  0.020_{-0.001}^{+0.001}  &  0.022_{-0.001}^{+0.001}  &  0.019_{-0.001}^{+0.001}  &  0.021_{-0.001}^{+0.001}  &  0.021_{-0.001}^{+0.001}\\
\DRpn(\B{11})  &  -0.052_{-0.014}^{+0.012}  &  -0.072_{-0.006}^{+0.005}  &  -0.082_{-0.006}^{+0.005}  &  -0.082_{-0.006}^{+0.006}  &  -0.068_{-0.001}^{+0.002}  &  -0.074_{-0.005}^{+0.005}  &  -0.077_{-0.004}^{+0.003}  &  -0.077_{-0.004}^{+0.003}\\
\DRpn(\B{12})  &  -0.146_{-0.013}^{+0.010}  &  -0.186_{-0.007}^{+0.009}  &  -0.190_{-0.007}^{+0.010}  &  -0.189_{-0.008}^{+0.011}  &  -0.167_{-0.008}^{+0.007}  &  -0.183_{-0.007}^{+0.007}  &  -0.188_{-0.009}^{+0.009}  &  -0.187_{-0.009}^{+0.009}\\[0.4em]
\DSRprms(\B{8,10})  &  1.786_{-0.109}^{+0.120}  &  1.523_{-0.115}^{+0.105}  &  1.627_{-0.080}^{+0.073}  &  1.620_{-0.094}^{+0.073}  &  1.893_{-0.111}^{+0.111}  &  1.474_{-0.081}^{+0.074}  &  1.630_{-0.064}^{+0.054}  &  1.630_{-0.079}^{+0.069}\\
\DSRprms(\B{8,11})  &  1.980_{-0.167}^{+0.127}  &  1.864_{-0.157}^{+0.120}  &  1.976_{-0.101}^{+0.092}  &  1.967_{-0.118}^{+0.083}  &  2.159_{-0.128}^{+0.128}  &  1.741_{-0.084}^{+0.077}  &  1.881_{-0.069}^{+0.053}  &  1.881_{-0.086}^{+0.069}\\
\DSRprms(\B{12,10})  &  -0.483_{-0.064}^{+0.048}  &  -0.185_{-0.061}^{+0.056}  &  -0.282_{-0.075}^{+0.057}  &  -0.282_{-0.081}^{+0.069}  &  -0.567_{-0.053}^{+0.048}  &  -0.231_{-0.054}^{+0.045}  &  -0.361_{-0.051}^{+0.046}  &  -0.361_{-0.064}^{+0.059}\\
\DSRprms(\B{12,11})  &  -0.282_{-0.023}^{+0.023}  &  0.178_{-0.060}^{+0.033}  &  0.080_{-0.056}^{+0.031}  &  0.072_{-0.068}^{+0.052}  &  -0.306_{-0.029}^{+0.024}  &  0.040_{-0.030}^{+0.023}  &  -0.112_{-0.023}^{+0.021}  &  -0.114_{-0.046}^{+0.044}
        \end{tabular}
    \end{ruledtabular}
    \caption{Numerical values of radius observables discussed in this work for the EMN interactions. Extrapolated results are given for each chiral order with many-body uncertainties and for the highest order with combined many-body and interaction uncertainties (see text for details).\label{tab:summary1}}
\end{table*}

\begin{table*}
    \centering
    \renewcommand{\arraystretch}{1.25}
    \begin{ruledtabular}
        \begin{tabular}{@{}l@{\extracolsep{13pt}}  d{2.8}@{\extracolsep{8pt}} d{2.8} d{2.8} d{2.7}@{\extracolsep{20pt}}   d{2.8}@{\extracolsep{8pt}} d{2.8} d{2.8} d{2.7}@{}}
            & \multicolumn{4}{c}{semilocal (SMS) $\Lambda=\qty{450}{\MeV}$} & \multicolumn{4}{c}{semilocal (SMS) $\Lambda=\qty{500}{\MeV}$}\\
        \cline{2-5}\cline{6-9}\\[-1.3em]
        Observable  &  \multicolumn{1}{c}{\NxLO{1}}  &  \multicolumn{1}{c}{\NxLO{2}}  &  \multicolumn{1}{c}{\NxLO[+]{4}}  &  \multicolumn{1}{c}{$+\chi$ uncert.} & \multicolumn{1}{c}{\NxLO{1}}  &  \multicolumn{1}{c}{\NxLO{2}}  &  \multicolumn{1}{c}{\NxLO[+]{4}}  &  \multicolumn{1}{c}{$+\chi$ uncert.}\\%
        \hline
\Rprms(\He{6})  &  1.818_{-0.011}^{+0.010}  &  1.793_{-0.007}^{+0.007}  &  1.844_{-0.008}^{+0.009}  &  1.844_{-0.024}^{+0.024}  &  1.857_{-0.009}^{+0.012}  &  1.776_{-0.007}^{+0.007}  &  1.827_{-0.008}^{+0.008}  &  1.827_{-0.028}^{+0.028}\\
\Rprms(\Li{8})  &  2.155_{-0.017}^{+0.015}  &  2.077_{-0.012}^{+0.013}  &  2.183_{-0.017}^{+0.013}  &  2.183_{-0.039}^{+0.038}  &  2.205_{-0.018}^{+0.019}  &  2.048_{-0.011}^{+0.013}  &  2.150_{-0.013}^{+0.014}  &  2.152_{-0.046}^{+0.046}\\
\Rprms(\B{8})  &  2.686_{-0.034}^{+0.029}  &  2.525_{-0.022}^{+0.024}  &  2.657_{-0.028}^{+0.026}  &  2.660_{-0.056}^{+0.060}  &  2.740_{-0.037}^{+0.037}  &  2.469_{-0.019}^{+0.021}  &  2.603_{-0.022}^{+0.024}  &  2.609_{-0.072}^{+0.070}\\
\Rprms(\B{10})  &  2.271_{-0.019}^{+0.026}  &  2.148_{-0.014}^{+0.016}  &  2.278_{-0.016}^{+0.019}  &  2.281_{-0.045}^{+0.048}  &  2.374_{-0.023}^{+0.025}  &  2.105_{-0.012}^{+0.012}  &  2.233_{-0.012}^{+0.015}  &  2.238_{-0.066}^{+0.066}\\
\Rprms(\B{11})  &  2.231_{-0.016}^{+0.020}  &  2.059_{-0.008}^{+0.012}  &  2.217_{-0.015}^{+0.017}  &  2.221_{-0.055}^{+0.055}  &  2.315_{-0.020}^{+0.022}  &  2.022_{-0.008}^{+0.010}  &  2.166_{-0.013}^{+0.015}  &  2.170_{-0.070}^{+0.071}\\
\Rprms(\B{12})  &  2.160_{-0.015}^{+0.016}  &  2.043_{-0.008}^{+0.010}  &  2.206_{-0.014}^{+0.014}  &  2.209_{-0.048}^{+0.049}  &  2.240_{-0.015}^{+0.020}  &  2.013_{-0.007}^{+0.009}  &  2.164_{-0.013}^{+0.013}  &  2.166_{-0.059}^{+0.062}\\[0.4em]
\DRpn(\He{6})  &  -0.880_{-0.032}^{+0.027}  &  -0.795_{-0.024}^{+0.019}  &  -0.848_{-0.025}^{+0.025}  &  -0.848_{-0.033}^{+0.033}  &  -0.867_{-0.030}^{+0.025}  &  -0.751_{-0.022}^{+0.017}  &  -0.804_{-0.023}^{+0.021}  &  -0.804_{-0.038}^{+0.036}\\
\DRpn(\Li{8})  &  -0.405_{-0.019}^{+0.018}  &  -0.349_{-0.010}^{+0.010}  &  -0.380_{-0.013}^{+0.012}  &  -0.381_{-0.019}^{+0.019}  &  -0.436_{-0.018}^{+0.018}  &  -0.348_{-0.014}^{+0.012}  &  -0.364_{-0.011}^{+0.010}  &  -0.365_{-0.023}^{+0.021}\\
\DRpn(\B{8})  &  0.492_{-0.020}^{+0.020}  &  0.420_{-0.011}^{+0.013}  &  0.458_{-0.014}^{+0.014}  &  0.461_{-0.022}^{+0.023}  &  0.493_{-0.018}^{+0.019}  &  0.396_{-0.009}^{+0.011}  &  0.424_{-0.016}^{+0.019}  &  0.425_{-0.026}^{+0.029}\\
\DRpn(\B{10})  &  0.023_{-0.002}^{+0.002}  &  0.020_{-0.001}^{+0.001}  &  0.022_{-0.001}^{+0.002}  &  0.022_{-0.002}^{+0.002}  &  0.025_{-0.002}^{+0.002}  &  0.019_{-0.001}^{+0.001}  &  0.007_{-0.013}^{+0.011}  &  0.007_{-0.013}^{+0.012}\\
\DRpn(\B{11})  &  -0.046_{-0.017}^{+0.014}  &  -0.065_{-0.003}^{+0.003}  &  -0.076_{-0.003}^{+0.003}  &  -0.076_{-0.005}^{+0.005}  &  -0.073_{-0.002}^{+0.003}  &  -0.064_{-0.003}^{+0.003}  &  -0.073_{-0.003}^{+0.003}  &  -0.073_{-0.004}^{+0.004}\\
\DRpn(\B{12})  &  -0.167_{-0.012}^{+0.010}  &  -0.187_{-0.009}^{+0.009}  &  -0.195_{-0.007}^{+0.007}  &  -0.194_{-0.009}^{+0.008}  &  -0.196_{-0.009}^{+0.011}  &  -0.180_{-0.010}^{+0.009}  &  -0.193_{-0.006}^{+0.007}  &  -0.192_{-0.008}^{+0.008}\\[0.4em]
\DSRprms(\B{8,10})  &  2.021_{-0.140}^{+0.140}  &  1.755_{-0.096}^{+0.081}  &  1.857_{-0.094}^{+0.103}  &  1.848_{-0.130}^{+0.112}  &  1.863_{-0.118}^{+0.118}  &  1.661_{-0.091}^{+0.064}  &  1.780_{-0.098}^{+0.083}  &  1.780_{-0.114}^{+0.098}\\
\DSRprms(\B{8,11})  &  2.218_{-0.130}^{+0.130}  &  2.144_{-0.099}^{+0.118}  &  2.127_{-0.115}^{+0.097}  &  2.127_{-0.115}^{+0.106}  &  2.148_{-0.125}^{+0.148}  &  2.003_{-0.100}^{+0.091}  &  2.076_{-0.103}^{+0.072}  &  2.076_{-0.103}^{+0.087}\\
\DSRprms(\B{12,10})  &  -0.498_{-0.070}^{+0.050}  &  -0.439_{-0.044}^{+0.040}  &  -0.326_{-0.039}^{+0.039}  &  -0.326_{-0.052}^{+0.046}  &  -0.602_{-0.067}^{+0.056}  &  -0.373_{-0.045}^{+0.037}  &  -0.316_{-0.033}^{+0.033}  &  -0.316_{-0.062}^{+0.059}\\
\DSRprms(\B{12,11})  &  -0.314_{-0.030}^{+0.030}  &  -0.068_{-0.010}^{+0.009}  &  -0.054_{-0.015}^{+0.015}  &  -0.055_{-0.051}^{+0.053}  &  -0.334_{-0.036}^{+0.036}  &  -0.037_{-0.010}^{+0.007}  &  -0.015_{-0.013}^{+0.011}  &  -0.017_{-0.061}^{+0.061}
        \end{tabular}
    \end{ruledtabular}
    \caption{Same as \cref{tab:summary1} but for the SMS interactions.\label{tab:summary2}}
\end{table*}

Lastly, we consider radius differences between nuclei.
As mentioned before, this observable is particularly noteworthy because it is experimentally accessible via isotope-shift measurements, e.g., in laser spectroscopy, allowing for direct comparison with theoretical predictions.
\Cref{fig:DSCRprms} presents the results for \DSRprms between the different boron isotopes.
Since \B{10} and \B{11} are stable and have already been measured experimentally, we express the differences with respect to either of these reference isotopes.
Note that none of the corresponding isotope shifts have been measured yet, however, experiments are in preparation \cite{Ber17,Ber19}.
For \B{8} we find a strongly increased radius with respect to both reference nuclei, which is consistent with a halo structure.
The differences are largely consistent across the interaction with a tendency to larger shifts for the SMS interactions.
Our results for \B{12} predict a shift to a smaller radius compared to \B{10}, while there seems to be no difference compared to \B{11}.
Again, the uncertainties for \DSRprms are reduced compared to a classical error propagation based on the \Rprms of the individual isotopes.

A collection of numerical values for the results discussed in this section is given in \cref{tab:summary1,tab:summary2}.

\section{Conclusions}

We have performed a precision NCSM study of radii in Boron isotopes enabled by our machine learning extrapolation tool that allows for combined many-body and interaction uncertainties.
Our results show that the point-proton radii are well reproduced for both EMN and SMS interactions up to a remaining cutoff dependence. The precision can further be increased through a direct treatment of radius differences, which also reduces interaction dependencies leading to remarkably stable predictions for the difference between point-proton and point-neutron radii, which are consistent with a proton halo in \B{8}. The same method can be applied to radius differences between different isotopes, which directly relate to experimentally accessible isotope shifts.
This paves the way for the comparison of theory predictions with upcoming measurements of isotope shifts in Boron isotopes.

\section*{Acknowledgments}

This work is supported by the Deutsche\linebreak Forschungsgemeinschaft (DFG, German Research Foundation) through the DFG Sonderforschungsbereich SFB 1245 (Project No.\ 279384907),
the BMBF through Verbundprojekt 05P2024 (ErUM-FSP T07, Contract No. 05P24RDB),
and by the U.S. Department of Energy, Office of Science, under Award No.\ DE-SC0023495 (SciDAC5/NUCLEI).
The NCSM calculations for the results presented here were performed with the code MFDn \cite{ccpe-10-2013-Aktulga,SHAO20181}
on Theta at the Argonne Leadership Facility (ALCF), a DOE Office of Science User Facility supported by the Office of Science of the U.S. Department of Energy under Contract DE-AC02-06CH11357, under the Innovative and Novel Computational Impact on Theory and Experiment (INCITE) program;
and on Perlmutter at the National Energy Research Scientific Computing Center (NERSC), a DOE Office of Science User Facility supported by the Office of Science of the U.S. Department of Energy under Contract No. DE-AC02-05CH11231, using NERSC Award No.\ NP-ERCAP0028672.
Numerical calculations for the training data have been performed on the LICHTENBERG II cluster at the computing center of the TU Darmstadt.

\bibliographystyle{apsrev4-2}
\bibliography{references.bib}

\end{document}